\documentclass[copyright,creativecommons,english]{eptcs}
\providecommand{\event}{DCM 2011} 
\usepackage{latexsym,amssymb,amsmath,mathdots,ellipsis,multirow,xcolor}
\usepackage{breakurl}     
\usepackage{algorithm}
\DeclareMathOperator{\sgn}{sgn}

\newcommand{\St}{\ensuremath{\mathcal {S}}}
\newcommand{\In}{\ensuremath{\St_0}}
\newcommand{\Ot}{\ensuremath{\St_\ddagger}}
\newcommand{\I}{\ensuremath{\mathcal{I}}}
\newcommand{\U}{\ensuremath{\mathcal {F}}}

\newcommand{\llbracket}{\left[\!\left[}
\newcommand{\rrbracket}{\right]\!\right]}
\newcommand{\evaluation}[2][]{\ensuremath{\llbracket #2\rrbracket_{#1}}}
\newcommand{\val}[2]{\evaluation[#2]{#1}}
\newcommand{\Dom}{\mathrm{Dom}~}
\renewcommand{\emptyset}{\varnothing}
\newcommand{\If}{\ensuremath{\mbox{\bf if}\;}}
\newcommand{\Then}{\;\mbox{\bf then}\;}
\newcommand{\Else}{\;\mbox{\bf else}\;}
\newcommand{\True}{\textsf{true}}
\newcommand{\False}{\textsf{false}}
\newcommand{\Undef}{\textsf{undef}}
\newcommand{\Do}{\;\textbf{do}\;}

\let\Del=\Delta
\renewcommand{\Delta}{{\rm \Del}}
\newtheorem{postulate}{Postulate}

\newtheorem{theorem}{Theorem}
\newtheorem{definition}[theorem]{Definition}
\renewenvironment{cases}{\left\{\begin{array}{ll}}{\end{array}\right.}
\title{The Generic Model of Computation}
\def\titlerunning{Generic Model of Computation}
\def\authorrunning{N. Dershowitz}
\author{Nachum Dershowitz
\institute{School of Computer Science\\Tel Aviv University\\Tel Aviv, Israel}
\email{\url{nachum.dershowitz@cs.tau.ac.il}}}
\date{\today}
\begin{document}
\maketitle

\begin{abstract}
Over the past two decades, Yuri Gurevich and his colleagues have formulated axiomatic foundations for the notion of \emph{algorithm}, be it
classical, interactive, or parallel, and formalized them in the new generic framework of \emph{abstract state machines}. 
This approach has recently been
extended to suggest a formalization of the notion of \emph{effective} computation over arbitrary countable domains.
The central notions are summarized herein.
\end{abstract}

\section{Background}

\textit{Abstract state machines (ASMs)}, invented by Yuri Gurevich
\cite{Gurevich94b}, constitute a most general model of computation,
one that can operate on any desired level of abstraction of data structures and native operations.
All (ordinary) models of computation are instances of this one generic paradigm.
Here, we give an overview of the foundational considerations underlying the model
(cobbled together primarily from~\cite{CT,Exact,Three}).%
\footnote{For a video lecture of Gurevich's on this subject, see~\url{http://www.youtube.com/v/7XfA5EhH7Bc}.}

Programs (of the sequential, non-interactive variety) in this formalism are built from three components:
\begin{itemize}
\item There are generalized assignments
$f(s_1,\ldots,s_n) := t$,
where $f$ is any function symbol  (in the vocabulary of the program)
and
the $s_i$ and $t$ are arbitrary terms (in that vocabulary).
\item
Statements may be prefaced by a conditional test,
$\If C \Then P$ or $\If C \Then P \Else Q$,
where $C$ is a propositional combination of equalities between terms.
\item Program statements may be composed in parallel, following the keyword
\textbf{do}, short for \textbf{do in parallel}.
\end{itemize}
An ASM program describes a single transition step; its statements are executed repeatedly, as a unit, until no assignments have their conditions enabled.
(Additional constructs beyond these are needed for interaction and large-scale parallelism,
which are not dealt with here.)

As a simple example, consider the program shown as Algorithm~\ref{sort}, describing a version of selection sort,
where $F(0),\dots,F(n-1)$ contain values to be sorted,
$F$ being a unary function symbol.
\begin{algorithm}[t]
\[
\renewcommand\arraystretch{1.4}
\If j = n 
\begin{array}[t]{ll}
\Then \If i+1 \neq n \Then \Do \begin{cases}
i := i+1 \\ j := i+2
 \end{cases}\\
\Else \Do \begin{cases}
\If F(i)>F(j) \Then \Do \begin{cases}
F(i) := F(j)  \\ F(j) := F(i) \end{cases}\\
 j := j+1
\end{cases}
\end{array}
\]
\caption{An abstract-state-machine program for sorting.}\label{sort}
\end{algorithm}
Initially,
$n\geq 1$ is the quantity of values to be sorted, $i$ is set to $0$, and $j$ to $1$.
The brackets indicate statements that are executed in parallel.
The program proceeds by repeatedly modifying the values of $i$ and $j$, as well as of locations
in $F$, referring to terms $F(i)$ and $F(j)$.
When all conditions fail, that is, when $j=n$ and $i+1=n$, 
the values in $F$
have been sorted vis-\`a-vis the black-box relation ``$>$''.
The program halts, as there is nothing left to do.
(Declarations and initializations for program constants and variables
are not shown.)

This sorting program is not partial to any particular representation of the natural numbers 1, 2, etc., which are being used to index $F$.
Whether an implementation uses natural language, or decimal numbers, or binary strings is immaterial, as long as addition
behaves as expected (and equality and disequality, too).
Furthermore, the program will work regardless of the domain from which the values of $F$ are drawn (be they integers, reals, strings, or what not), so long as means are provided for evaluating
the inequality ($>$) relation.

\begin{algorithm}[t]
\[
\renewcommand\arraystretch{1.4}
\If |b-a|>\varepsilon \Then \Do 
\begin{cases}
\If \sgn f((a+b)/2) = \sgn f(a) \Then a := (a+b)/2\\ 
\If \sgn f((a+b)/2) = \sgn f(b) \Then b := (a+b)/2
\end{cases}
\]
\caption{An abstract-state-machine program for bisection search.}\label{cacm}
\end{algorithm}

Another simple ASM program is shown in Algorithm~\ref{cacm}.
This is a standard bisection search for the root of a function,
as described in~\cite[Algorithm \#4]{cacm:alg4}.
The point is that this abstract formulation is, as the author of \cite{cacm:alg4} wrote, ``applicable to any continuous function'' over the reals---including ones that cannot be programmed.

What is remarkable about ASMs is that this very simple model of computation suffices
to precisely capture the behavior of the whole class of ordinary algorithms over any domain.
The reason is that, by virtue of the abstract state machine (ASM) representation theorem of~\cite{Gurevich00} (Theorem~\ref{thm:asm} below),
any algorithm that satisfies three very natural ``Sequential Postulates''
can be \textit{step-by-step, state-for-state} emulated by an ASM.
Those postulates, articulated in Section~\ref{sec:ax}, formalize the following intuitions: (I) an algorithm is a state-transition system;
(II) given the algorithm, state information determines future transitions and can be captured by a logical structure;
and (III) state transitions are governed by the values of a finite and input-independent set of terms.

The significance of the Sequential Postulates lies in their comprehensiveness.
They formalize which features exactly characterize a classical algorithm in its most
abstract and generic manifestation.
Programs of all models of effective, sequential computation satisfy the postulates, as
do idealized algorithms for computing with real numbers (e.g.\ Algorithm~\ref{cacm}), or for geometric constructions
with compass and straightedge
(see~\cite{Reisig04} for examples of the latter).

Abstract state machines are a computational model that
is not wedded to any particular data representation, in the way, say,
that Turing machines manipulate strings using a small set of tape operations.
The Representation Theorem, restated in Section~\ref{sec:asm}, establishes that ASMs can express
and precisely emulate any and all algorithms satisfying the premises captured by the postulates.
For any such algorithm, there is an ASM program that describes precisely the
same state-transition function, state after state, as does the algorithm.
In this sense, ASMs subsume all other computational models.

It may be informative to note the similarity between the form of an ASM, namely, a single repeated loop of a set of generalized assignments nested within conditionals with the ``folk theorem'' to the effect that any flowchart program can be converted to a single loop composed of conditionals, sequencing, and assignments,
with the aid of some auxiliary variables (see \cite{Harel}).
Parallel composition gives ASMs the ability to perform multiple actions sans extra variables, and to capture all that transpires in a single step of any algorithm.

This versatility of ASMs is what makes them so ideal for both specification and prototyping.
Indeed, ASMs have been used to model all manner of programming applications, systems,
and languages, each on the precise intended level of abstraction. See~\cite{Boerger02a} 
and the ASM website (\url{http://www.eecs.umich.edu/gasm}) for numerous exemplars. 
ASMs provide a complete means of describing algorithms, whether or not they can be implemented effectively.
On account of their abstractness, one can express generic algorithms, like our bisection search for arbitrary continuous real-valued functions, or like
Gaussian elimination, even when the field over which it is applied
is left unspecified.
AsmL~\cite{AsmL}, an executable specification
language based on the ASM framework, has been used in industry, in particular
for the behavioral specification of interfaces (see, for example,~\cite{BarSch01b}).

Church's Thesis asserts that the recursive functions are the only numeric functions that can be effectively computed. Similarly, Turing's
Thesis stakes the claim that any function on strings that can be mechanically computed can be computed, in particular, by a Turing machine. 
More generally, one additional natural hypothesis regarding the
describability of initial states of algorithms,
as explained in Section~\ref{sec:eff},
characterizes the effectiveness
of any model of computation, operating over any (countable) data domain
(Theorem~\ref{thm:ctt}).

On account of the ability of ASMs to precisely capture single steps of any algorithm, one can infer absolute bounds on the complexity of algorithms under arbitrary effective models of computation, as will be seen (Theorem~\ref{thm:ectt}) at the end of Section~\ref{sec:eff}.

\section{Sequential Algorithms}\label{sec:ax}

The Sequential Postulates of~\cite{Gurevich00}
regarding algorithmic behavior are based on the following key observations:
\begin{itemize}
\item A state should contain \emph{all} the relevant information, apart from the algorithm itself, needed to determine the next steps.
For example, the ``instantaneous description'' of a Turing machine computation is just what is needed to pick up a machine's computation
from where it has been left off; see~\cite{Turing}.
Similarly, the ``continuation'' of a Lisp program contains all the state information needed to resume its computation.
First-order structures suffice to model all salient features of states. Compare~\cite[pp.~420--429]{Post}.
\item The values of programming variables, in and of themselves, are meaningless to an
algorithm, which is implementation independent.  
Rather, it is relationships between values that matter to the algorithm.
It follows that an algorithm should work equally well in isomorphic worlds.
Compare \cite[p.~128]{Gandy}.
An algorithm can---indeed, can only---determine relations between values stored in a state via terms in its vocabulary
and equalities (and disequalities) between their values.
\item Algorithms are expressed by means of finite texts, making reference to only finitely many terms and relations among them.
See, for example, \cite[p.~493]{Kleene87}.
\end{itemize}

The three postulates given below (from~\cite{Gurevich00}, modified slightly as in~\cite{Ord1,Ord2,Ord3,Exact}) assert that a classical algorithm is a state-transition system operating over first-order structures in a way that is invariant under isomorphisms.
An algorithm is a prescription for updating states, that is, for changing some of the interpretations given to symbols by states.
The essential idea is that there is a fixed finite set of terms that refer (possibly indirectly) to locations within a state
and which suffice to determine how the state changes during any transition.

\subsection{Sequential Time}

To begin with,
algorithms are deterministic state-transition systems.

\begin{postulate}[Sequential Time]\label{P1}
An algorithm determines the following:
\begin{itemize}
\item A nonempty set\/\footnote{Or class; the distinction is irrelevant for our purposes.}
$\St$ of \emph{states} and a nonempty subset $\In \subseteq \St$ of \emph{initial} states.\item A partial \emph{next-state} transition function $\tau:\St\rightharpoonup\St$.
\end{itemize}
\end{postulate}

\emph{Terminal} states $\Ot \subseteq \St$ are those states $X$
for which no transition $\tau(X)$ is defined.

Having the transition depend only on the state
means that states must store all the information needed to determine subsequent behavior.
Prior history is unavailable to the algorithm unless stored in the current state.

State-transitions are deterministic.
Classical algorithms in fact never leave room
for choices, nor do they involve any sort of interaction with the environment to determine
the next step.
To incorporate nondeterministic choice, probabilistic choice, or interaction with the environment, one
would need to modify the above notion of transition.

This postulate is meant to exclude formalisms, such as~\cite{Gold,Putnam}, 
in which the result of a computation---or the continuation of a computation---may depend
on (the limit of) an infinite sequence of preceding (finite or infinitesimal) steps.
Likewise,
processes in which states evolve continuously (as in analog processes, like the position of a bouncing ball), rather than discretely, are eschewed.

Though it may appear at first glance that a recursive function does not fit under the rubric of a state-transition system,
in fact the definition of a traditional recursive function comes together with
a computation rule for evaluating it.
As Rogers~\cite[p.~7]{Rogers} writes,
``We obtain the computation uniquely by working from the inside out and
from left to right''.

\subsection{Abstract State}

Algorithm states are comprehensive:
they incorporate all the relevant data (including any ``program counter'')
that, when coupled with the program, completely determine the future of a computation.
States may be regarded as structures with (finitely many) functions, relations, and constants.
To simplify matters, relations will be treated as truth-valued functions
and constants as nullary functions.
So, each state consists of a domain (base set, universe, carrier) and interpretations for its symbols.
All relevant information about a state
is given explicitly in the state by means of its interpretation of the symbols
appearing in the vocabulary of the structure.
The specific details of the implementation of the data types used by the algorithm cannot matter.
In this sense states are ``abstract''. This crucial consideration leads to the second postulate.

\begin{postulate}[Abstract State]\label{P2}
The states $\St$ of an algorithm are (first-order) structures
over a finite vocabulary $\U$, such that the following hold:
\begin{itemize}
\item If $X$ is a state of the algorithm, then any structure $Y$ that is isomorphic to $X$ is also a state, and $Y$ is initial or terminal if $X$ is initial or terminal, respectively.
\item Transitions preserve the domain; that is, $\Dom\tau(X)=\Dom X$ for every non-terminal state $X$.
\item Transitions respect isomorphisms, so, if $\zeta:X\cong Y$ is an isomorphism of non-terminal states $X,Y$, then also $\zeta: \tau(X)\cong\tau(Y)$.
\end{itemize}
\end{postulate}

State structures are endowed with
Boolean truth values and standard Boolean operations, and vocabularies include symbols for these.
As a structure, a state interprets each of the function symbols in its vocabulary.
For every $k$-ary symbol
$f$ in the vocabulary of a state $X$ and values
 $a_1,\dots,a_k$ in its domain,
some domain value $b$
is assigned to the \emph{location} $f(a_1,\dots,a_k)$, for which
we write  $f(\bar a)\mapsto b$.
In this way, $X$ assigns a value $\val{t}{X}$ in $\Dom X$ to (ground) terms $t$.

Vocabularies are finite, since an algorithm must be describable in finite terms,
so can only refer explicitly to finitely many operations.
Hence, an algorithm can not, for instance, involve all of Knuth's arrow operations,
$\uparrow$, 
$\uparrow
\uparrow$, 
$\uparrow
\uparrow 
\uparrow$, 
etc.
Instead one could employ a ternary operation $\lambda x,y,z.\; x\uparrow^z y$.

This postulate is justified by the vast experience of mathematicians and scientists who have faithfully and transparently
presented every kind of static mathematical or scientific reality as a logical structure.

In restricting structures to be ``first-order'', we are limiting the \emph{syntax} to be first-order.
This precludes states with infinitary operations, like the supremum of infinitely many objects, which would not make sense from an algorithmic point of view.  This does not, however, limit the semantics of algorithms to first-order notions.  The domain of states may have sequences, or sets, or other higher-order objects, in which case, the state would also need to provide operations for dealing with those objects.

Closure under isomorphism ensures that the algorithm can
operate on the chosen level of abstraction.
The states' internal representation of data is invisible and immaterial to the program.
This means that the behavior of an \textit{algorithm}, in contradistinction with its ``implementation'' as a C program---cannot,
for example, depend on the memory address of some variable. 
If an algorithm does depend on such matters, then its full description must also include specifics of memory allocation.

It is possible to liberalize this postulate somewhat to allow the domain to grow or shrink, or for the vocabulary to be infinite
or extensible,
but such ``enhancements'' do not materially change the notion of algorithm.
An extension to structures with partial operations is given in \cite{Exact}; see Section~\ref{sec:thm}.

\subsection{Effective Transitions}

The actions taken by a transition are describable in terms of updates
of the form $f(\bar a)\mapsto b$, meaning that $b$ is the \emph{new} interpretation to be given by the next state
to the function symbol $f$ for values $\bar a$.
To program such an update, one can use an assignment $f(\bar s):=t$ such that $\val{\bar s}{X}=\bar a$ and $\val{t}{X}=b$.
We view a state $X$  as a collection of the graphs of its operations,
each point of which is a location-value pair also denoted $f(\bar a)\mapsto b$.
Thus, we can define the \emph{update set} $\Delta(X)$ as the changed points, $\tau(X)\setminus X$.
When $X$ is a terminal state and $\tau(X)$ is undefined, we indicate that by setting
$\Delta(X)=\bot$.

The point is that $\Delta$ encapsulates the state-transition relation $\tau$ of an algorithm
by providing all the information necessary to update the interpretation given by the current state.
But to produce $\Delta(X)$ for a particular state $X$, the algorithm needs to evaluate some terms  with the help of the information stored in  $X$.
The next postulate will ensure that $\Delta$ has a finite representation and its updates can be determined and performed
by means of only a finite amount of work.
Simply stated, there is a fixed, finite set of ground terms that determines the stepwise behavior of an algorithm.

\begin{postulate}[Effective Transitions]\hspace{-5pt}\footnote{Or \textbf{Bounded Exploration}.}\label{BE}
For every algorithm, there is a finite set $T$ of (ground) \emph{critical terms}  over the state 
 vocabulary, such that
states that agree on the values of the terms in $T$ also share the same update sets.
That is,
$ \Delta(X)=\Delta(Y)$,
for any two states $X,Y$
such that $\val t X = \val t Y$ for all $t\in T$.
In particular, if one of $X$ and $Y$ is terminal, so is the other.
\end{postulate}

The intuition is that an algorithm must base its actions on the
values contained at locations in the current state.
Unless all states undergo the same updates unconditionally, an algorithm must
explore one or more values at some accessible locations in the current state before determining how to proceed.
The only means that an algorithm has with which to reference locations is via terms,
since the values themselves are abstract entities.
If every referenced location has the same value in two states,
then the behavior of the algorithm must be the same for both of those states.

This postulate---with its fixed, finite set of critical terms---precludes programs of infinite size (like an infinite table lookup) or which 
are input-dependent.

A careful analysis of the notion of algorithm in~\cite{Gurevich00}
and an examination of the intent of the founders of the field of computability in~\cite{CT}
demonstrate that the Sequential Postulates are in fact
true of all ordinary, sequential algorithms, the (only) kind envisioned
by the pioneers of the field.
In other words, all \textit{classical} algorithms satisfy Postulates \ref{P1}, \ref{P2}, and \ref{BE}.
In this sense, the traditional notion of algorithm is precisely captured by these axioms.

\begin{definition}[Classical Algorithm]\label{def:class}
An object satisfying Postulates \ref{P1}, \ref{P2}, and \ref{BE} shall be called a \emph{classical algorithm}.
\end{definition}

\subsection{Equivalent Algorithms}

It makes sense to say that two algorithms have the same behavior, or are \emph{behaviorally equivalent}, if they operate over the same states and have the same transition function.

Two algorithms are \emph{syntactically equivalent} if their states are the same up to renaming of symbols ($\alpha$-conversion)
in their vocabularies, and if transitions are the same after renaming.

For a wide-ranging discussion of algorithm equivalence, see~\cite{same}.

\section{Abstract State Machines}\label{sec:asm}

Abstract state machines (ASMs) are an all-powerful description language for the classical algorithms we have been characterizing.

\subsection{Programs}

The semantics of the ASM statements, assignment, parallel composition, and
conditionals, are as expected, and are formalized below.
The program, as such, defines a single step, which is repeated forever or until
there is no next state.

For convenience,
we show only a simple form of ASMs.
Bear in mind, however, that much
richer languages for ASMs are given in~\cite{Gurevich94b} and are used in practice
\cite{GuScVe01b}.

Programs are expressed in terms of some vocabulary.
By convention, ASM programs always include symbols for the Boolean values (\True\ and \False), \Undef\ for a default, ``undefined'' value, standard Boolean operations
($\neg$, $\wedge$, $\vee$), and equality ($=,\neq$).
The vocabulary of the sorting program, for instance, contains $\U=\{1,2,+,>,F,n,i,j\}$ in addition to the standard symbols.
Suppose that its states have integers and the three standard values for their domain.
The nullary symbols $0$ and $n$ are fixed programming constants and serve as bounds of $F$.
The nullary symbols $i$ and $j$ are programming ``variables'' and are used as array indices.
All its states interpret the symbols $1,2,+,>$, as well as the standard symbols, as usual.
Unlike $i$, $j$, and $F$, these are static; their
interpretation will never be changed by the program.
Initial states have $n\geq 0$, $i=0$, $j=1$, some integer values for $F(0),\dots,F(n-1)$,
plus \Undef\ for all other points of $F$.
This program always terminates successfully, with $j=n=i+1$
and with the first $n$ elements of $F$ in nondecreasing order.

There are no hidden variables in ASMs.
If some steps of an algorithm are intended to be executed in sequence, say, then
the ASM will need to keep explicit track of where in the sequence it is up to.

\subsection{Semantics}\label{sec:up}

Unlike algorithms, which are observed to
either change the value of a location in the current state, or not,
an ASM might ``update'' a location in a \emph{trivial} way, giving it the same value it already has.
Also, an ASM might designate two conflicting updates for the same location, what is called a \emph{clash}, in which case
the standard ASM semantics are to cause the run to fail (just as real-world programs might abort).
An alternative semantics is to imagine a nondeterministic choice between the competing values.
(Both were considered in~\cite{Gurevich94b}.)
Here, we prefer to ignore both nondeterminism and implicit failure, and 
tacitly presume that an ASM never involves clashes, albeit this
is an undecidable property.

To take the various possibilities into account, a \emph{proposed} update set $\Delta^+_P(X)$ (cf.~\cite{Ord1}) for an ASM $P$ may be defined in the following manner:
\[
\renewcommand\arraystretch{1.3}
\begin{array}{rcl}
\Delta^+_{f(s_1,\dots,s_n) := t}(X) &=& \{f(\val{s_1}X,\dots,\val{s_n}X) \mapsto  \val tX\}\\[2pt]
\Delta^+_{\mbox{\bf\small do } \{P_1 \cdots P_n\}}(X) &=& 
\Delta^+_{P_1}(X)\cup\cdots\cup \Delta^+_{P_n}(X) 
\\
\Delta^+_{\mbox{\bf\small if } C \mbox{ \bf \small then } P \mbox{ \bf \small else } Q}(X) &=&
\begin{cases}
\Delta^+_P(X) & \mbox{if $X\models C$}\\
\Delta^+_Q(X)  & \mbox{otherwise}
\end{cases}\\
\Delta^+_{\mbox{\bf\small if } C \mbox{ \bf \small then }P}(X) &=&
\begin{cases}
\Delta^+_P(X) & \mbox{if $X\models C$}\\
\emptyset & \mbox{otherwise}\;.
\end{cases}
\end{array}
\]
Here $X\models C$ means, of course, that Boolean condition $C$ holds true in $X$.
When the condition $C$ of a conditional statement does not evaluate to $\True$,
the statement does not contribute any updates.

When $\Delta^+(X)=\emptyset$ for ASM $P$, its execution halts with success,
in terminal state $X$.
(Since no confusion will arise, we are dropping the subscript $P$.)
Otherwise, the updates are applied to $X$ to yield the next state
by replacing the values of all locations in $X$ that are referred to in $\Delta^+(X)$.
So, if the latter
contains only trivial updates, $P$ will loop forever.

For terminal states $X$, the update set $\Delta(X)$ is $\bot$, to signify that there is no next state.
For non-terminal $X$, $\Delta(X)$ is the set of non-trivial updates in $\Delta^+(X)$.
The update sets for the  sorting program (Algorithm~\ref{sort}) are shown in Table~\ref{tab},
with the subscript in $\val{\cdot}X$ omitted.
\begin{table}
\renewcommand\arraystretch{1.4}
\[
\begin{array}{|l||c|c|}\hline
  & \mbox{States $X$ such that} & \mbox{Update set $\Delta(X)$}\\\hline\hline
0 & \val{j}{}=\val{n}{}=\val{i}{}+1 & \bot\\\hline
1 & \val{j}{}=\val{n}{}\neq\val{i}{}+1 & i\mapsto \val{i}{}+1 ,\;  j\mapsto \val{i}{}+2\\\hline
2 & \val{j}{}\neq\val{n}{},\;\val{F(i)}{}>\val{F(j)}{} & 
F(\val{i}{})\mapsto \val{F(j)}{},\;F(\val{j}{})\mapsto \val{F(i)}{},  j\mapsto \val{j}{}+1
\\\hline
3 & \val{j}{}\neq\val{n}{},\;\val{F(i)}{}\not> \val{F(j)}{} & j\mapsto \val{j}{}+1\\\hline
\end{array}
\]
\caption{Update sets for sorting program.}\label{tab}
\end{table}
For example, if state $X$ is such that  $n=2$, $i=0$, $j=1$, $F(0)=1$, and $F(1)=0$,
then (per row 2)
$\Delta^+(X)=\{F(0)\mapsto 0, F(1)\mapsto 1, j\mapsto 2\}$.
For this $X$,
$\Delta(X)=\Delta^+(X)$, and
the next state $X'=\tau(X)$ has
$i=0$ (as before), $j=2$, $F(0)=0$ and $F(1)=1$.
After one more step (per row 1), in which $F$ is unchanged,
the algorithm reaches a terminal state, $X''=\tau(X')$, with $j=n=i+1=2$.
Then (by row 0), $\Delta^+(X'')=\emptyset$ and $\Delta(X'')=\bot$.

\section{The Representation Theorem}\label{sec:thm}

Abstract state machines clearly satisfy the three Sequential Postulates:
ASMs define a state-transition function; they operate over abstract states;
and they depend critically on the values of a finite set of terms appearing in the program
(and on the unchanging values of parts of the state not modified by the program).
For example, the critical terms for our sorting ASM are all the terms appearing in it,
except for the left-hand sides of assignments, which contribute their proper subterms instead.
These are
$ j\neq n$, $(j = n) \wedge (i+1 \neq n)$, $F(i)>F(j)$, $i+2$, $j+1$,
and their subterms.
Only the values of these affect the computation.
Thus, any ASM describes a classical algorithm over structures with the same vocabulary (similarity type).

The converse is of greater significance:

\begin{theorem}[Representation {\cite[Theorem 6.13]{Gurevich00}}]\label{thm:asm}
Every classical algorithm, in the sense of Definition~\ref{def:class},
has a behaviorally equivalent ASM,
with the exact same states and state-transition function.
\end{theorem}
The proof of this representation theorem constructs an ASM that contains conditions involving equalities and disequalities between critical terms.
Closure under isomorphisms is an essential ingredient for making it possible to
express any algorithm in the language of terms.

A typical ASM models partial functions (like division or tangent) by using the special value, \Undef, denoting
that the argument is outside the function's domain of definition, and arranging
that most operations be strict, so a term involving an undefined subterm is likewise undefined.
The state of such an ASM would return \True\ when asked to evaluate
an expression $c/0 = \Undef$, and it can, therefore, be programmed to work properly,
despite the partiality of division.

In~\cite{Exact}, the analysis and representation theorem have been refined for algorithms employing truly partial operations, 
operations that cause an algorithm to hang when an operation is attempted outside its domain of definition
(rather than return \Undef\/).
The point is that 
there is a behaviorally equivalent ASM that never attempts to access locations in the state that are not also accessed by the given algorithm.
Such partial operations are required in the next section.

\section{Effective Algorithms}\label{sec:eff}

The Church-Turing Thesis~\cite[Thesis I$^\dagger$]{Kleene67} asserts that standard models capture effective computation.
Specifically:
\begin{quote}
All effectively computable numeric (partial) functions are (partial) recursive.\\
All (partial) string functions can be computed by a Turing machine.
\end{quote}

We say that an algorithm \emph{computes} a partial function
$f:D^k\rightharpoonup D$ if there are \emph{input} states $\I\subseteq\In$, with particular locations for input values, such that
running the algorithm results in the correct output values of $f$.
Specifically:
\begin{itemize}
\item The domain of each input state is $D$.  There are $k$ terms such that their values in input states cover all tuples in $D^k$.  
Other than that, input states all agree on the values of all other terms.
\item For all  input values $\bar a$, the corresponding input state leads,
via a sequence of transitions $\tau$,
to a terminal state in which the value of a designated term $t$ (in the vocabulary of the algorithm) is $f(\bar a)$ whenever the latter is defined, and leads to an infinite computation whenever it is not.
\end{itemize}

To capture what it is that makes a sequential algorithm mechanically computable, we  need for
input states to be finitely representable.
Accordingly, we insist that they harbor no information beyond the means to reach domain values, plus anything that can be derived therefrom.

We say that function symbols $\cal C$ \emph{construct} domain $D$ in state $X$
if $X$ assigns each value in $D$ to exactly one term over $\cal C$,
so restricting $X$ to $\cal C$ gives a free Herbrand algebra.
For example, the domain of the sorting algorithm, consisting of integers and Booleans, can be constructed from $0,\True,\False,\Undef$, and a 
``successor'' function (call it $c$) that takes non-negative integers ($n$) to the predecessor of their negation ($-n-1$) and negative integers ($-n$) to their absolute value ($n$).

Postulate~\ref{BE} ensures that the transition function is describable by a finite text, and---in particular--by the text of ASM.
For an algorithm to be effective, its states must also be finitely describable.

\begin{definition}[Effectiveness]\ 
\begin{enumerate}
\item A state is \emph{effective} if it includes
constructors for its domain,
plus operations that are almost everywhere the same,
meaning that all but finitely-many locations (these can hold input values) have the same default value (such as \Undef\/).
\item A classical algorithm is \emph{effective} if its initial states are.
\item Moreover, effective algorithms can be bootstrapped: 
A state is effective also if its vocabulary can be enriched to
$\cal C\uplus \cal G$ so that $\cal C$ constructs its domain, while
every (total or partial) operation in $\cal G $ is computed by an effective algorithm over those constructors.
\item A \emph{model (of computation)}, that is, a set of algorithms with shared domain(s), is \emph{effective} if all its algorithms are, via the \emph{same} constructors.
\end{enumerate}
\end{definition}

This effectiveness postulate excludes algorithms with ineffective oracles, such as the halting function.
Having only free constructors at the foundation precludes the hiding of potentially uncomputable information by means of equalities between distinct representations of the same domain element.

This is the approach to effectiveness advocated in~\cite{CTT}, extended
to include partial functions in states, as in~\cite{Exact}.
For each $n\geq 1$, our sorting algorithm is effective in this sense,
since addition ($+$) of the natural numbers and comparisons ($>$) of integers, operations that reside in its initial states, can be programmed from the above-mentioned constructors ($0,\True,\False,\Undef,c$).

In particular, partial-recursion for natural numbers and Turing machines for strings form effective models~\cite{CTT}.
Furthermore, it is shown in~\cite{Three}
that three prima facie different definitions of effectiveness over arbitrary domains, as proposed in~\cite{CTT,CT,ComputableKernel}, respectively,
comprise exactly the same functions,
strengthening the conviction that the essence of the underlying notion of computability has in fact been captured.

\begin{theorem}[{Church-Turing Thesis~\cite{CTT}}]\label{thm:ctt}
For every effective model, there is a representation of its domain values
as strings, such that its algorithms are each simulated by some Turing machine.
\end{theorem}

Call an effective computational model 
\emph{maximal} if adding any function to those that it computes results in a
set of functions that cannot be simulated by any effective model. 
Remarkably (or perhaps not), there is exactly one such model:

\begin{theorem}[{Effectiveness~\cite[Theorem 4]{Three}}]
The set of partial recursive functions (and likewise the set of Turing-computable string functions) is the unique maximal effective model, up to isomorphism, over any countable domain.
\end{theorem}

We have recently extended the proof of the Church-Turing Thesis and
demonstrated the validity of the widely believed \textit{Extended Church-Turing Thesis}: 

\begin{theorem}[Extended Church-Turing Thesis \cite{ECCT}]\label{thm:ectt}
Every effective
algorithm can be polynomially simulated by a Turing machine. 
\end{theorem}

\section{Conclusion}\label{sec:end}

We have dealt herein with the classical type of algorithms, that is to say, with the ``small-step'' (meaning, only bounded parallelism) ``sequential-time'' (deterministic, no intra-step interaction with the
outside world) case.
Abstract state machines can faithfully emulate any algorithm in this class, as we have seen in Theorem~\ref{thm:asm}.
Furthermore, we have characterized the distinction between effective algorithms and their more abstract siblings in Theorem~\ref{thm:ctt}.

There are various ``declarative'' styles of programming for which the state-transition relation is implicit, rather than explicit as it is for our notion of algorithm.
For such programs to be algorithms in the sense of Definition~\ref{def:class}, they would have to be equipped with a specific execution mechanism, like the one for recursion mentioned above.
For Prolog, for example, the mechanism of unification and the mode of search would need to be specified~\cite{Prolog}.

The abstract-state-machine paradigm can be extended to handle more modern notions:
\begin{itemize}
\item
When desired, an algorithm can make an explicit distinction between successful and failing
terminal states by storing particular values
in specific locations of the final state.
Alternatively, one may declare failure when there is a conflict between two or more enabled assignments. See \cite{Gurevich94b}.
\item
There is no difficulty in allowing for nondeterminism, that is, for a multivalued transition function.
If the semantics are such that a choice is made between clashing assignment statements, then transitions are indeed nondeterministic.
See \cite{Gurevich94b,GY}.
\item
More general forms of nondeterminism can be obtained by adding a choice command of some sort to the language.
See~\cite{Gurevich94b}.
\item Nothing needs to be added to the syntax of ASMs to apply to cases for the environment provides input incrementally.
One need only imagine that the environment is allowed to modify the values of
some (specified) set of locations in the state between machine steps. See \cite{Gurevich94b}.
\item
In~\cite{Ord1,Ord2,Ord3}, 
the analysis of algorithms was extended to the case when an algorithm interacts with the outside environment during a step,
and execution waits until all queries of the environment have been responded to.
\item In \cite{General1,General2}, all forms of interaction are handled.
\item In \cite{parallel},
the analysis was extended to massively parallel algorithms.
\item
Distributed algorithms are handled in~\cite{Gurevich94b,Glausch}.
\item
The fact that ASMs can emulate algorithms step-for-step facilitates reasoning
about the complexity of algorithms, as for Theorem~\ref{thm:ectt} above.
Parallel ASMs have been used for studying the complexity of algorithms over unordered structures. See \cite{BGS,Spielmann}.
\item
Quantum algorithms have been modeled by ASMs in~\cite{Quantum}.
\item
Current research includes an extension of the framework for hybrid systems,
combining discrete (sequential steps) and analog (evolving over time) behaviors
\cite{Analog,TAMC}.
\end{itemize}

\subsection*{Acknowledgements}

I thank Yuri Gurevich and Nikolaj Bj{\o}rner  for their perspicacious suggestions,
the referees for their questions,
and Evgenia Falkovich for her help.

\bibliographystyle{eptcs}
\bibliography{models}
\end{document}